\documentclass[%
superscriptaddress,
preprintnumbers,
tightenlines,
showpacs,
showkeys,
nofootinbib,
amsmath,
amssymb,
aps,
prd,
floatfix,
]{revtex4-1}

\usepackage{graphicx}
\usepackage{color}

\newcommand{\dof}{{\rm d.o.f.}}
\newcommand{\be}{\begin{equation}}
\newcommand{\ee}{\end{equation}}
\newcommand{\bea}{\begin{eqnarray}}
\newcommand{\eea}{\end{eqnarray}}

\begin{document}
\title{Nature of the phase transition for finite temperature $N_{\rm f}=3$ QCD\\
  with nonperturbatively O($a$) improved Wilson fermions at $N_{\rm t}=12$}

\author{Yoshinobu Kuramashi}
\affiliation{Center for Computational Sciences, University of Tsukuba, Tsukuba, Ibaraki 305-8577, Japan}

\author{Yoshifumi~Nakamura}
\affiliation{RIKEN Center for Computational Science, Kobe 650-0047, Japan}

\author{Hiroshi Ohno}
\affiliation{Center for Computational Sciences, University of Tsukuba, Tsukuba, Ibaraki 305-8577, Japan}

\author{Shinji Takeda}\email[]{takeda@hep.s.kanazawa-u.ac.jp}
\affiliation{Institute of Physics, Kanazawa University, Kanazawa 920-1192, Japan}

\date{\today}
\begin{abstract}
We study the nature of the finite temperature phase transition for three-flavor QCD.
In particular we investigate the location of the critical endpoint along the three flavor symmetric line in the light quark mass region of the Columbia plot.
In the study, the Iwasaki gauge action and the nonperturvatively O($a$) improved Wilson-Clover fermion action are employed.
We newly generate data at $N_{\rm t}=12$
and set an upper bound of the critical pseudoscalar meson mass in the continuum limit
$m_{\rm PS,E}\lesssim 110$~MeV.
\end{abstract}


\preprint{UTHEP-744, UTCCS-P-129, KANAZAWA-20-01}

\maketitle

\section{Introduction}
\begin{figure}[hbt!]
\begin{center}
\includegraphics[bb= 0 0 739 642, width=0.49\textwidth]{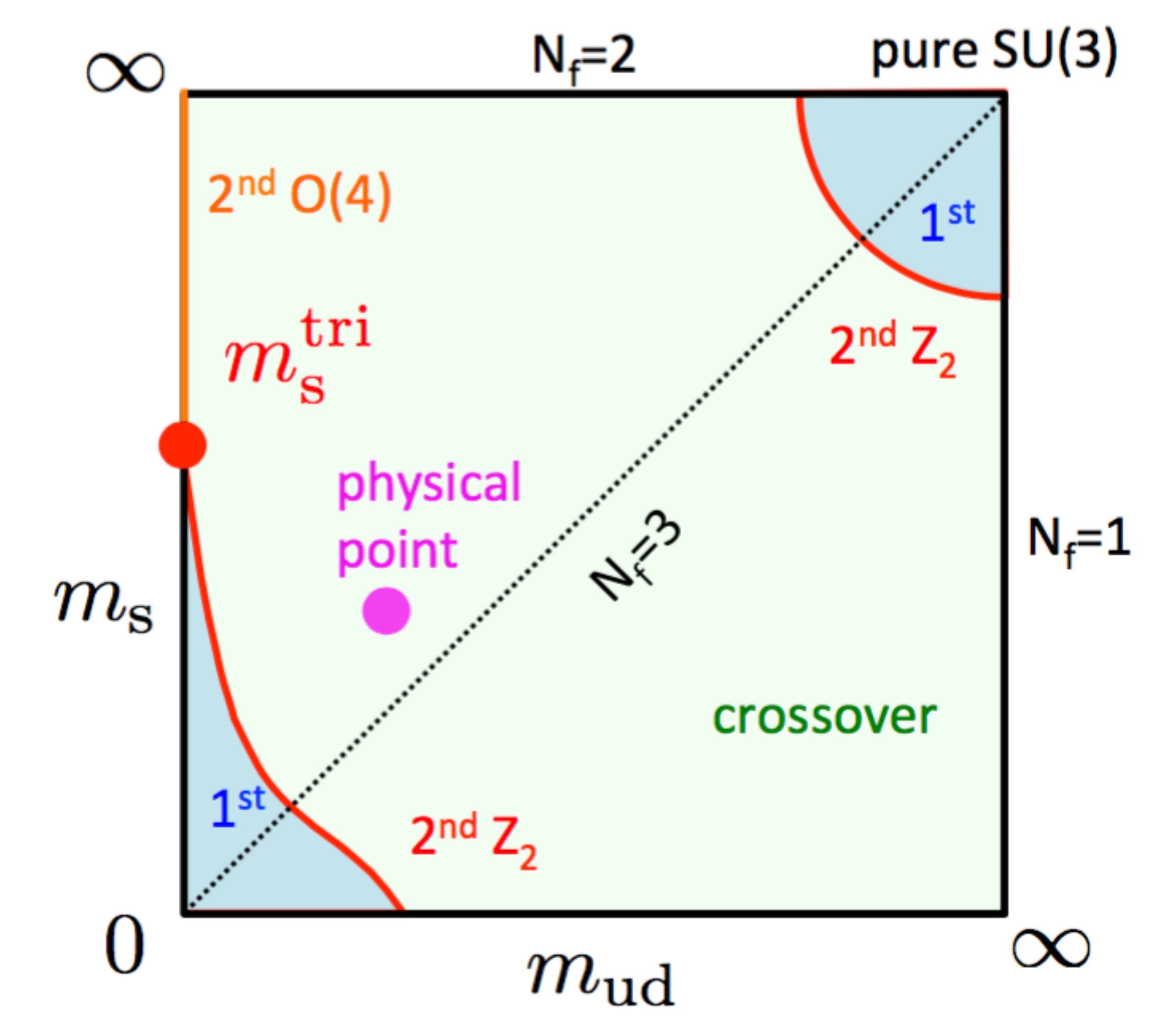}
\end{center}
\caption{Columbia plot for $N_{\rm f}=2+1$ QCD at zero density.
}
\label{fig:ColumbiaPlot}
\end{figure}

The finite temperature transition in QCD is an important subject in elementary particle physics and cosmology.
The nature of the finite temperature phase transition has been studied over a number of years.
So far there are some analytic attempts to investigate the nature of the phase transition;
effective theories based on the universality argument~\cite{Pisarski:1983ms,Gavin:1993yk,Butti:2003nu,Calabrese:2004uk} were systematically studied
and recently an anomaly matching argument~\cite{Shimizu:2017asf,Yonekura:2019vyz} has been developed.
Although these approaches can capture qualitative aspects,
it is hard to provide its quantitative information on the nature of the phase transition
without fully taking the nonperturbative effects of QCD.
Lattice QCD simulations play central roles in revealing the quantitative aspects, and in fact
many efforts have been devoted for this aim.
See reviews~\cite{Schmidt:2017bjt,Ding:2017giu,Sharma:2019wiv,Philipsen:2019rjq}
for a current status of the QCD phase structure with the finite temperature and quark number density.
In such studies, the so-called Columbia plot~\cite{Brown:1990ev} is often used
to express the nature of the phase transition in various parameter space.
A whole structure of the plot is basically dictated by a critical point, line or surface, which separates the first order and crossover region,
depending on the dimensionality of the parameter space.
It is therefore crucial to figure out the shape of such critical boundaries.
The standard Columbia plot in the case of zero density has two axes: the up-down and strange quark masses (See Fig.~\ref{fig:ColumbiaPlot}).
In the heavy mass region of the plot, especially the static limit is well established as the first order phase transition~\cite{Brown:1988qe,Fukugita:1989yb} and
the heavy region apart from the static limit is also studied~\cite{Saito:2011fs,Czaban:2016yae}.
On the other hand, the light quark mass region is still under debate and
we will closely investigate such region in the following.
Although there are interesting issues for two-flavor QCD, for example restoration of the U$_A$(1) symmetry and so on
(see Refs.~\cite{Schmidt:2017bjt,Ding:2017giu,Sharma:2019wiv,Philipsen:2019rjq} for recent progress),
in this paper we restrict ourselves to the three-flavor symmetric case where all three quark masses are degenerated.
In particular, our goal is to locate the critical endpoint along the flavor-symmetric line on the standard Columbia plot.

Let us look back on a historical background of the location of the critical endpoint for the three-flavor QCD.
A rough but first estimate of the critical endpoint was provided by Iwasaki {\it et al.},~\cite{Iwasaki:1996zt} using the Wilson-type fermions and their
critical quark mass is relatively heavy $m_{\rm q,E}\gtrsim140$~MeV, equivalently $m_{\rm PS,E}\gtrsim O(1)$~GeV in terms of the pseudoscalar meson mass.
Subsequently a study with the standard staggered fermion action was carried out by JLQCD collaboration~\cite{Aoki:1998gia} 
and they estimated $am_{\rm q,E}\approx0.03$ using screening mass analysis.
A similar study was done by Liao~\cite{Liao:2001en} and similar conclusion was drawn.
Then Karsch {\it et al.},~\cite{Karsch:2001nf} reported $m_{\rm PS,E}\approx290$~MeV using the Binder intersection method with the combination of the standard staggered fermions and the Wilson plaquette gauge action,
and in addition they also estimated $m_{\rm PS,E}\approx190$~MeV using an improved staggered-type fermion action, i.e., p4-action.
In Ref.~\cite{Karsch:2003va}, they updated $m_{\rm PS,E}=67(18)$~MeV for the p4-action with replacing the gauge action to the Symanzik-improved one,
then a large cutoff effect on the critical endpoint was indicated.
Although the R-algorithm~\cite{Gottlieb:1987mq} was used in the staggered fermion studies mentioned above, 
de Forcrand and Philipsen~\cite{deForcrand:2006pv} performed rational hybird Monte Carlo (RHMC) simulation~\cite{Clark:2004cp,Clark:2006fx} and
found $am_{\rm q,E}=0.0260(5)$ which is significantly smaller than the previous value
with the R-algorithm.
Smith and Schmidt~\cite{Smith:2011pm} examined the RHMC results using larger spatial volumes and it was confirmed that the critical point belongs to the three-dimensional Z$_2$ universality class.
In the above staggered studies, a single lattice spacing was exclusively used (the temporal lattice size was fixed to be $N_{\rm t}=4$), however,
de Forcrand {\it et al.},~\cite{deForcrand:2007rq} extended their study to see the lattice cutoff dependence and found that
$m_{\rm PS,E}/T_{\rm E}$, $T_{\rm E}$ is temperature at the critical endpoint, decreases from $1.680(4)$ to $0.954(12)$ as increasing $N_{\rm t}$ from $4$ to $6$.
This explicitly shows that it is important to control the cutoff effects on the critical point and also suggests that the critical mass in the continuum limit
may be quite small.
Further studies were continued using the improved staggered fermions with smearing techniques~\cite{Endrodi:2007gc,Ding:2011du,Varnhorst:2015lea,Bazavov:2017xul},
but they could not even detect a critical point.
Instead, for example, Ding {\it et al.},~\cite{Bazavov:2017xul} quoted an upper bound of the critical mass $m_{\rm PS,E}\lesssim50$~MeV.

In the above situation, we embarked a study of the nature of the phase transition using the $O(a)$ improved Wilson fermions instead of the staggered-type fermions.
Such a study is important to check the universality when taking the continuum limit
and our formulation is completely free of the rooting issue~\cite{Creutz:2007rk,Bernard:2007eh}.
In the early stage of our study~\cite{Jin:2014hea} with coarse lattice spacings $N_{\rm t}=4$, $6$ and a part of $N_{\rm t}=8$,
we observed a quite large scaling violation in the continuum extrapolation of the critical endpoint.
Therefore we extended our study to $N_{\rm t}=8$ and $10$~\cite{Jin:2017jjp} together with the multiensemble reweighting technique.
Then we confirmed the universality class of the critical endpoint to be Z$_2$ universality class for $N_{\rm t}=4$ and $6$,
while it is assumed for $N_{\rm t}=8$ and $10$ and we used a modified fitting form of the Binder (kurtosis) intersection analysis.
And then we set an upper bound $m_{\rm PS,E}<170$~MeV.
In the current paper, we further extend our study and generate the new data set of $N_{\rm t}=12$ in order to take the continuum limit smoothly
and make sharpe the prediction of the critical point if exists.

The rest of the paper is organized as follows.
We describe the simulation setup and the analysis methods in Sec.~\ref{sec:setup}. 
In Sec.~\ref{sec:results}, we locate the critical point by applying two analysis methods for a cross check.
Then we discuss the continuum limit of the critical pseudoscalar mass and the critical temperature.
Our conclusions are summarized in Sec.~\ref{sec:summary}.
Results of zero temperature simulations for scale setting are summarized in Appendix~\ref{sec:scale}.

\section{Setup and methods}
\label{sec:setup}

Our finite temperature $N_{\rm f}=3$ QCD simulations are performed with the Iwasaki gauge action~\cite{Iwasaki:2011np} and
nonperturvatively O($a$) improved Wilson-Clover fermion action~\cite{Aoki:2005et}.
In this paper we report our newly generated data with the temporal lattice size of $N_{\rm t}=12$.
To carry out the finite size scaling analysis 5 different spatial lattice sizes with $N_{\rm s}=16$, $20$, $24$, $28$, and $32$ are used.
As we will see soon, the smaller spatial lattices $16$ and $20$ are used only when estimating the transition point in the thermodynamic limit
and the critical points will be determined using the larger volumes $24$, $28$, and $32$, which satisfy $m_{\rm PS} L\gtrsim4$.
Gauge configurations are generated with the RHMC algorithm~\cite{Clark:2006fx} implemented with the Berlin QCD code~\cite{Nakamura:2010qh},
where the acceptance rate is tuned to be around $70$--$80$\%. Observables are measured at every 10th molecular dynamics trajectory whose length is set to unity.
There is a single hopping parameter $\kappa$ for three degenerate dynamical flavors in our simulations, which is adjusted to search for
a transition point at each $\beta$, where $\beta$ values are chosen in a range between $1.80$ and $1.82$.
See Table~\ref{tab:ensemble} for the parameter sets and their statistics.

\begin{table}[!hbt]
\begin{ruledtabular}
\caption{\label{tab:ensemble}Simulation parameters and the number of configurations.}
\begin{tabular}{ll|rrrrr}
$ \beta $ & $\kappa$ & $N_{\rm s}=16$& $N_{\rm s}=20$& $N_{\rm s}=24$ & $N_{\rm s}=28$ & $N_{\rm s}=32$ \\
\hline
  $ 1.80 $&$ 0.139150 $&$ 1040 $&$    - $&$    - $&$    -$&$    -$ \\
  $      $&$ 0.139200 $&$ 1040 $&$  650 $&$    - $&$    -$&$    -$ \\
  $      $&$ 0.139220 $&$  830 $&$  900 $&$  300 $&$    -$&$    -$ \\
  $      $&$ 0.139240 $&$     - $&$ 700 $&$    - $&$    -$&$    -$ \\
  $      $&$ 0.139250 $&$ 1150 $&$    - $&$    - $&$    -$&$    -$ \\
\hline
  $ 1.81 $&$ 0.138950 $&$  730 $&$    - $&$    - $&$    -$&$    -$ \\
  $      $&$ 0.139000 $&$  660 $&$    - $&$    - $&$    -$&$    -$ \\
  $      $&$ 0.139020 $&$    - $&$ 1310 $&$    - $&$    -$&$    -$ \\
  $      $&$ 0.139040 $&$    - $&$    - $&$ 3850 $&$ 2660$&$  910$ \\
  $      $&$ 0.139050 $&$  700 $&$ 1540 $&$ 1200 $&$    -$&$ 1330$ \\
\hline
  $ 1.82 $&$ 0.138800 $&$  700 $&$    - $&$    - $&$    -$&$    -$ \\
  $      $&$ 0.138810 $&$    - $&$  560 $&$    - $&$    -$&$    -$ \\
  $      $&$ 0.138820 $&$    - $&$    - $&$    - $&$  960$&$ 1400$ \\
  $      $&$ 0.138830 $&$    - $&$  600 $&$ 4210 $&$ 2510$&$ 1550$ \\
  $      $&$ 0.138850 $&$    - $&$  910 $&$    - $&$    -$&$    -$ \\
  $      $&$ 0.138880 $&$  740 $&$    - $&$    - $&$    -$&$    -$ \\
\end{tabular}
\end{ruledtabular}
\end{table}

\begin{figure}[hbt!]
\begin{center}
\includegraphics[bb= 0 0 480 285, width=0.7\textwidth]{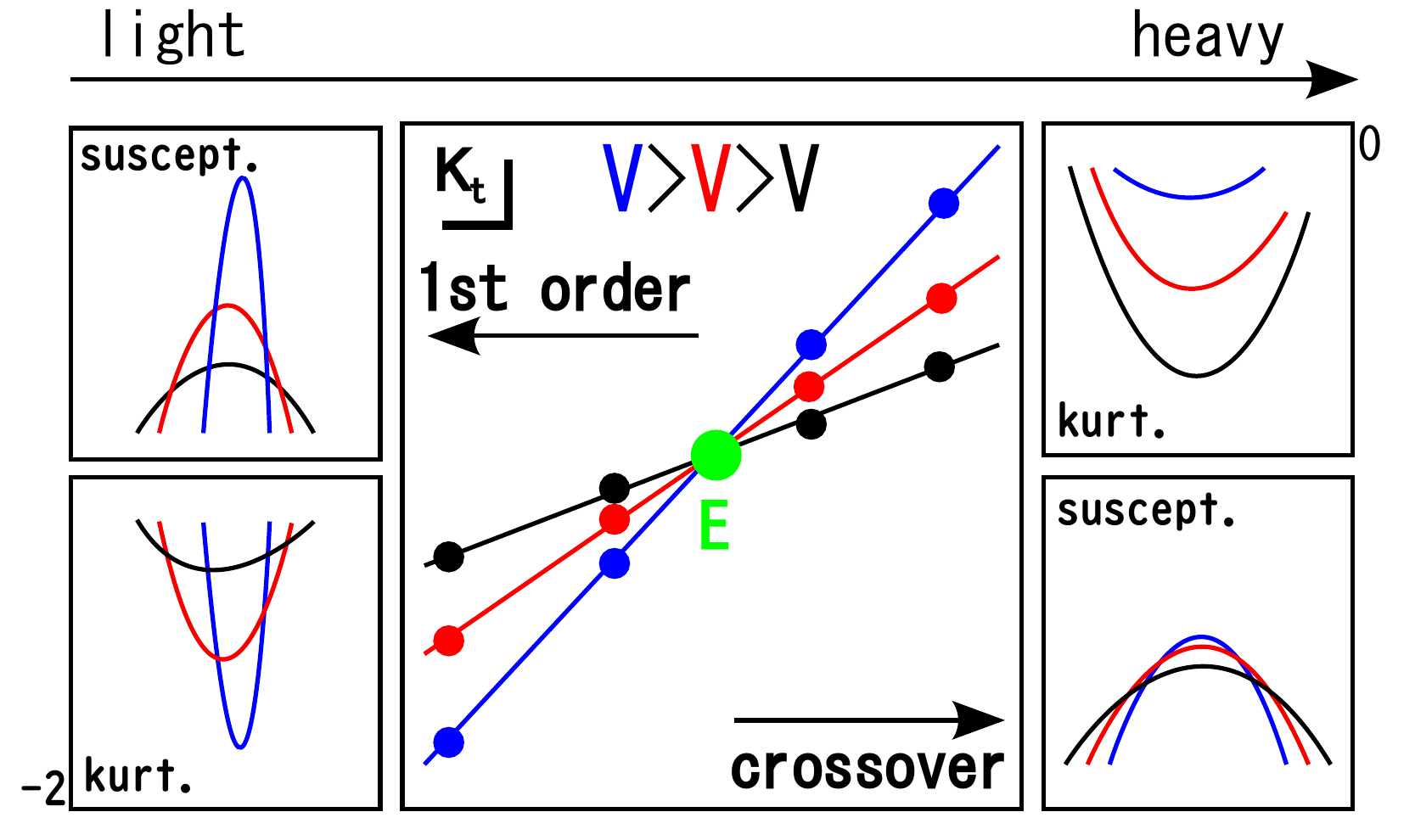}
\end{center}
\caption{An illustration of the kurtosis intersection analysis. $K_t$ denotes the kurtosis value for the phase transition and $E$ indicates the critical endpoint,
where $K_t$ is independent of the volume.
}
\label{fig:kurtosis}
\end{figure}

We follow the same analysis methods as our previous studies~\cite{Jin:2014hea,Kuramashi:2016kpb,Jin:2017jjp}, which are summarized in the following:

\begin{enumerate}
\item The chiral condensate and its higher order moments up to the fourth are measured. The definition of the moments is given in our previous paper~\cite{Kuramashi:2016kpb}.
\item The multiensemble reweighting~\cite{Ferrenberg:1988yz} in only $\kappa$ but not $\beta$ is used, which enables us to smoothly interpolate the moments. The reweighting factor, which is given by the ratio of fermion determinants at different $\kappa$ values, is calculated with an expansion of the logarithm of the determinant~\cite{Kuramashi:2016kpb}. Adopting an expansion form for the moments in the reweighting method, we can evaluate the moments at continuously many points at a relatively low cost.
\item From these moments the susceptibility, the skewness, and the kurtosis, which is equivalent to the Binder cumulant up to an additional constant, are calculated.
\item The $\kappa$ value at the transition point is estimated from the peak position of the susceptibility at each $\beta$.
\item After repeating the procedure 1--4 for a few spatial lattice sizes, the location of the critical point is estimated by the kurtosis intersection analysis~\cite{Karsch:2001nf}, where we search for a point at which the kurtosis value for the phase transition is independent of the volume as schematically illustrated in Fig.~\ref{fig:kurtosis}. In the determination of the critical point we use a fit ansatz with the inclusion of the energy-like observable contribution~\cite{Jin:2017jjp}.
\end{enumerate}

\section{Results}
\label{sec:results}

\subsection{Moments and location of the transition point}

\begin{figure}[t!]
\begin{center}
\includegraphics[bb= 0 0 274 244, width=0.49\textwidth]{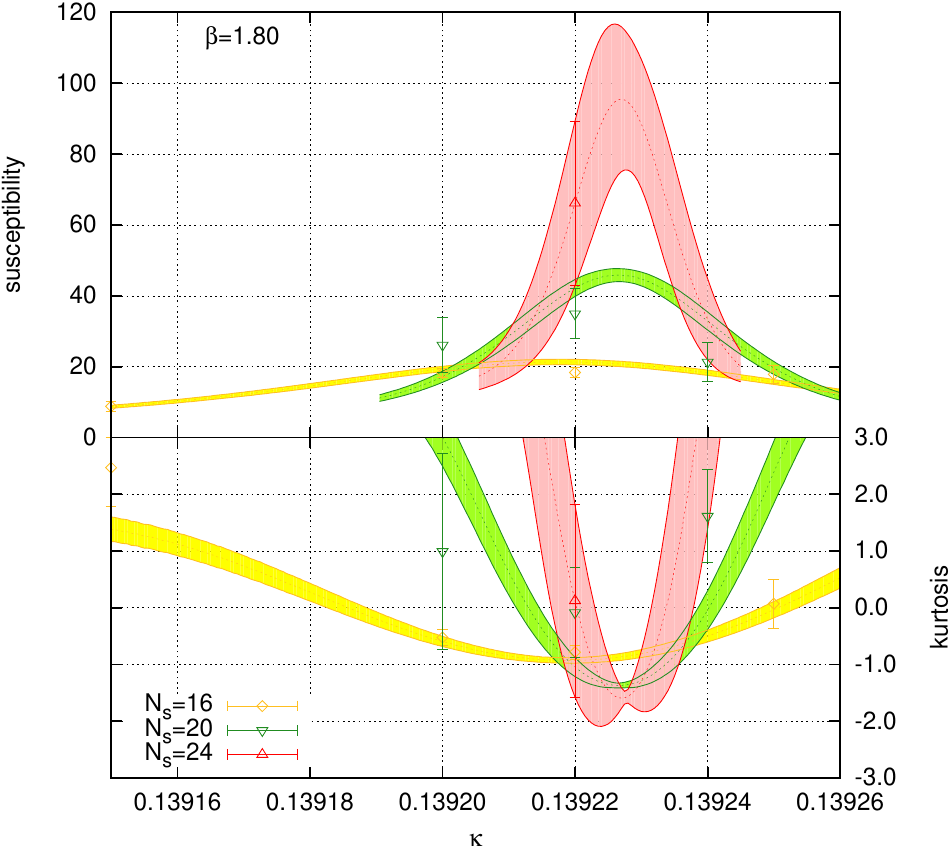}
\includegraphics[bb= 0 0 274 244, width=0.49\textwidth]{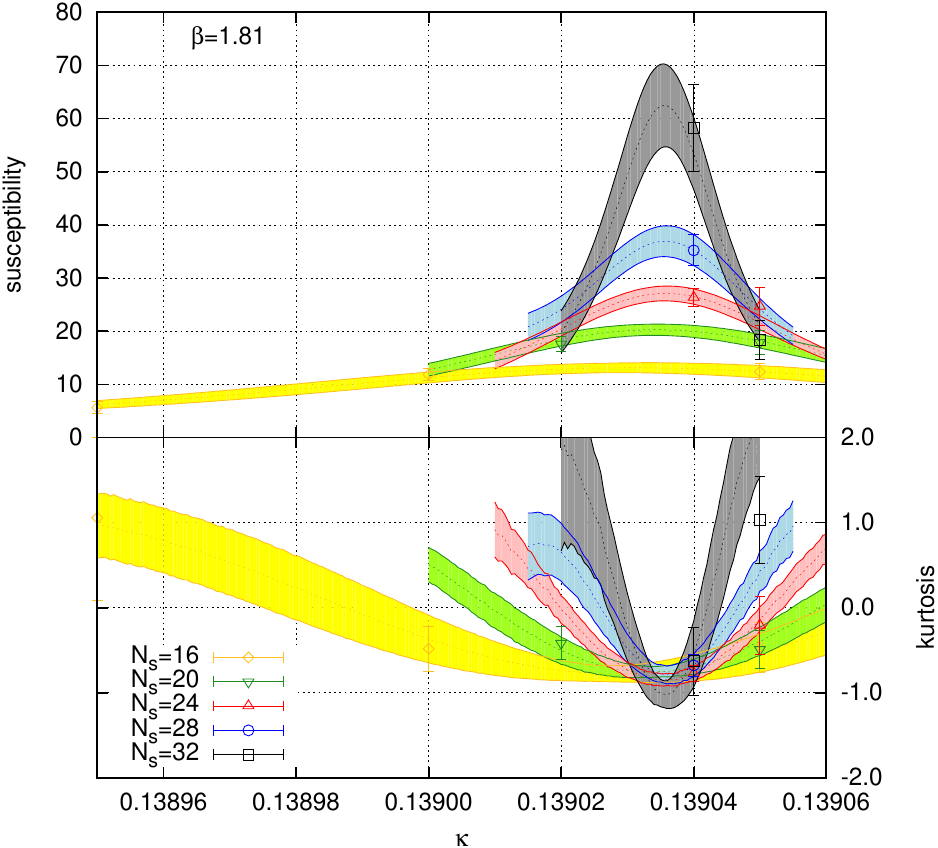}
\end{center}
\caption{The susceptibility (upper half) and kurtosis (lower half) of the chiral condensate as a function $\kappa$
for several spatial sizes, $N_{\rm s}=16$--$32$.
The left (right) panel is for $\beta=1.80$ ($\beta=1.81$).
The raw data points (as symbols) as well as the multiensemble reweighting results (1-$\sigma$ band) are plotted.
}
\label{fig:moment}
\end{figure}

\begin{figure}[h!]
\begin{center}
\includegraphics[bb= 0 0 315 221, width=0.7\textwidth]{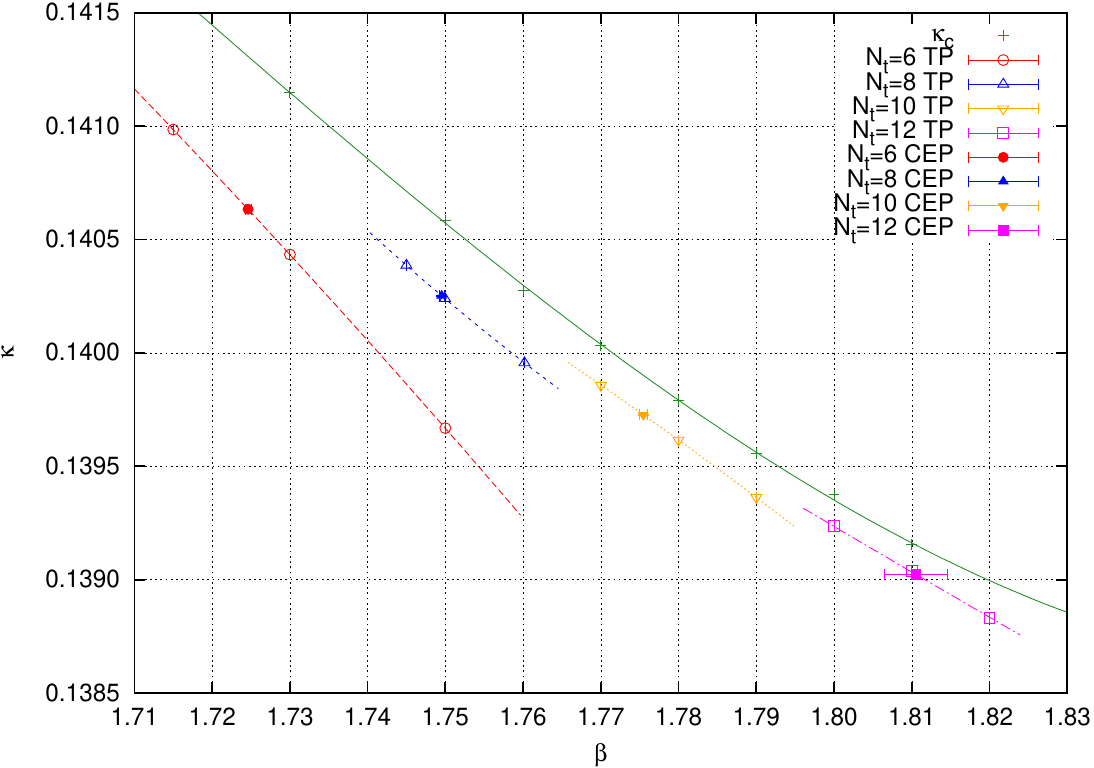}
\end{center}
\caption{
Phase diagram for the bare parameter space $(\beta,\kappa)$ at $N_{\rm t}=12$
together with $N_{\rm t}=6$, $8$, and $10$ results in Ref.~\cite{Jin:2017jjp}.
The open symbols represent a transition point (TP) while the
filled symbols are the critical endpoints (CEP)
determined by the kurtosis intersection with a correction term.
On the transition line, the left (right) hand side of an
critical endpoint
is the first order phase transition (crossover) side.
The phase transition line for each $N_{\rm t}$ is a polynomial interpolation.
For $N_{\rm t}=12$ the interpolation formula is given in Eq.~(\ref{eqn:kappa_t}).
In the plot, $\kappa_{\rm c}$ is the pseudoscalar massless point with $N_{\rm f}=3$ which is determined by zero temperature simulations as shown in Table~\ref{tab:scalefit} in Appendix~\ref{sec:scale}.
}
\label{fig:phase_diagram}
\end{figure}

As an illustration of the data, we show the susceptibility and the kurtosis of the chiral condensate
for $\beta=1.80$ and $1.81$ in Fig.~\ref{fig:moment} together with the $\kappa$-reweighting results.
From the peak position of the susceptibility, we extract the value of $\kappa$ at transition points denoted as $\kappa_{\rm t}(\beta,N_{\rm s})$,
whose values are summarized in Table~\ref{tab:kappat} for various $N_{\rm s}$ and $\beta$.
The peak height of the susceptibility and the minimum of the kurtosis are also shown in the table.
For each value of $\beta$, the infinite volume limit of the transition point $\kappa_{\rm t}(\beta,N_{\rm s}=\infty)$
is carried out by using a fitting form with an inverse spatial volume correction term\footnote{
There is no specific meaning in using $1/N_{\rm s}^3$ correction form.
As seen in Table~\ref{tab:kappat}, we do not observe a significant $N_{\rm s}$-dependence
on $\kappa_{\rm t}$, thus a choice of extrapolation form seems irrelevant.
In fact, we performed a fit with $\exp(-m_{\rm PS}L)$ correction term with fixed $m_{\rm PS}=0.2$
whose value is estimated from Table~\ref{tab:scale} with the corresponding $\beta$ and $\kappa$ value.
As a result, the thermodynamic value of $\kappa_{\rm t}$ is consistent with
that obtained with $1/N_{\rm s}^3$ correction form.
},
\be
\kappa_{\rm t}(\beta,N_{\rm s})=\kappa_{\rm t}(\beta,N_{\rm s}=\infty)+c(\beta)/N_{\rm s}^3\,,
\label{eqn:kappa_t_V}
\ee
where $\kappa_{\rm t}(\beta,N_{\rm s}=\infty)$ and $c(\beta)$ are fitting parameters.
For $\beta=1.80$, three smaller volumes $N_{\rm s}=16$, $20$, and $24$ are used while all five volumes are used for $\beta=1.81$ and $1.82$.
The quality of the fitting is reasonable with $\chi^2/\dof<1.7$ for all cases.
The resulting phase diagram in the bare parameter space is summarized in Fig.~\ref{fig:phase_diagram}.
The phase transition line in the thermodynamic limit (we denote $\kappa_{\rm t}(\beta)=\kappa_{\rm t}(\beta,\infty)$) at $N_{\rm t}=12$
is determined by the linear interpolation,
\be
\kappa_{\rm t}(\beta)
=
0.139238(4)-(\beta-1.8)\times0.0203(4)\,.
\label{eqn:kappa_t}
\ee

\begin{table}[!hbt]
\begin{ruledtabular}
\caption{\label{tab:kappat}
Summary of the value of $\kappa_{\rm t}$: the value of $\kappa$ at the transition point, $\chi_{\max}$: the maximum of susceptibility and $K_{\min}$: the minimum of kurtosis
for each $N_{\rm s}$ and $\beta$.
The thermodynamics limit of $\kappa_{\rm t}$ is taken with the fitting form in Eq.~(\ref{eqn:kappa_t_V}).
The errors of $\chi_{\max}$ and $K_{\min}$ are estimated by the jackknife analysis.
}
\begin{tabular}{ll|lll}
$\beta$
&
$N_{\rm s}$
&
$\kappa_{\rm t}(\beta,N_{\rm s})$
&
$\chi_{\max}$
&
$K_{\min}$
\\
\hline
$ 1.80 $
&$ 16 $&$ 0.1392177 ( 39 ) $&$ 21.3 ( 1.3 ) $&$ -0.931 ( 68 ) $\\
&$ 20 $&$ 0.1392265 ( 33 ) $&$ 45.8 ( 2.3 ) $&$ -1.373 ( 51 ) $\\
&$ 24 $&$ 0.1392270 ( 65 ) $&$ 95.5 ( 7.8 ) $&$ -1.595 ( 14 ) $\\
&$\infty$&$ 0.1392348 ( 54 ) $& & \\
\hline
$ 1.81 $
&$ 16 $&$ 0.1390307 ( 94 ) $&$ 13.12 ( 95 ) $&$ -0.78 ( 10 ) $\\
&$ 20 $&$ 0.1390343 ( 40 ) $&$ 20.3 ( 1.1 ) $&$ -0.761 ( 58 ) $\\
&$ 24 $&$ 0.1390358 ( 16 ) $&$ 27.1 ( 1.4 ) $&$ -0.850 ( 71 ) $\\
&$ 28 $&$ 0.1390357 ( 16 ) $&$ 36.9 ( 2.9 ) $&$ -0.79 ( 12 ) $\\
&$ 32 $&$ 0.1390356 ( 16 ) $&$ 62.5 ( 7.8 ) $&$ -1.01 ( 16 ) $\\
&$\infty$&$ 0.1390364 ( 19 ) $& & \\
\hline
$ 1.82 $
&$ 16 $&$ 0.138832 ( 15 ) $&$ \phantom{0}9.35 ( 58 ) $&$ -0.68 ( 10 ) $\\
&$ 20 $&$ 0.1388217 ( 57 ) $&$ 12.02 ( 72 ) $&$ -0.61 ( 14 ) $\\
&$ 24 $&$ 0.1388370 ( 44 ) $&$ 15.07 ( 78 ) $&$ -0.42 ( 18 ) $\\
&$ 28 $&$ 0.1388273 ( 41 ) $&$ 16.1 ( 1.1 ) $&$ -0.30 ( 10 ) $\\
&$ 32 $&$ 0.1388285 ( 46 ) $&$ 16.8 ( 1.0 ) $&$ -0.50 ( 16 ) $\\
&$\infty$&$ 0.1388305 ( 43 ) $& & \\
\end{tabular} 
\end{ruledtabular}
\end{table}

\subsection{Kurtosis intersection analysis}

\begin{figure}[h!]
\begin{center}
\includegraphics[bb= 0 0 246 168, width=0.49\textwidth]{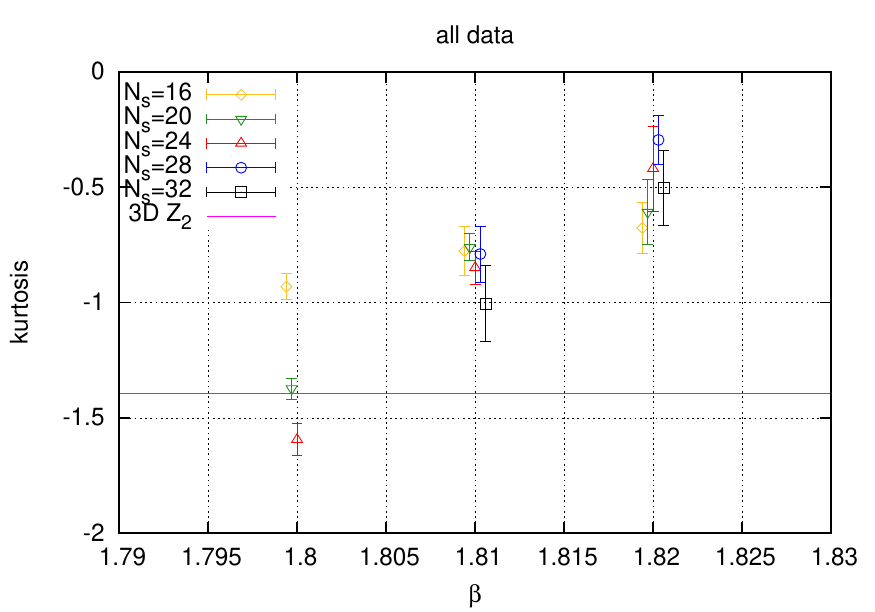}
\includegraphics[bb= 0 0 246 168, width=0.49\textwidth]{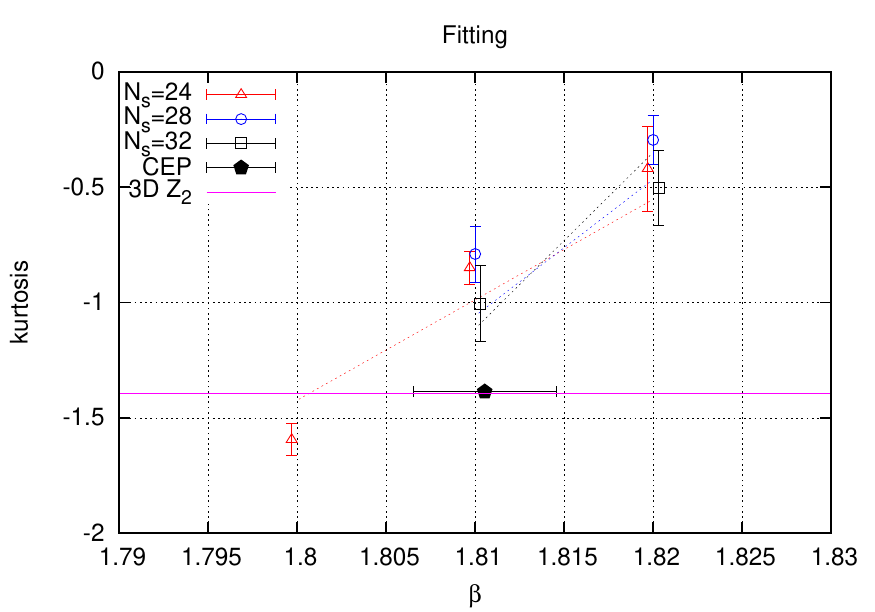}
\end{center}
\caption{
Kurtosis intersection for the chiral condensate at $N_{\rm t}=12$.
The left panel contains all data points of $N_{\rm s}=16-32$.
The right panel includes larger volumes $N_{\rm s}\ge24$ together with the fitting function in Eq.~(\ref{eqn:new}) which assumes the 3D Z$_2$ universality class
and contains a correction term.
The black pentagon represents the resulting critical value of $\beta$.
}
\label{fig:krt}
\end{figure}

The minimum of kurtosis is plotted in Fig.~\ref{fig:krt}
to perform kurtosis intersection analysis at $N_{\rm t}=12$.
The left panel of Fig.~\ref{fig:krt} includes all $N_{\rm s}$ data points.
At $\beta=1.80$, a typical behavior of the first order phase transition is clearly seen;
the kurtosis tends to be smaller with increasing the volume.
On the other hands the results at $\beta=1.81$ show volume independent behavior.
For $\beta=1.82$, apart from $N_{\rm s}=32$ data point which has a relatively larger error bar,
the crossover behavior is seen in the volume dependence; for larger volume the kurtosis tends to be larger.
Therefore it is likely that there is a crossing point between $\beta=1.80$ and $1.82$.
To keep away from the finite size effects we use $N_{\rm s}\ge24$ data points ($m_{\rm PS}L\ge4$)
for the kurtosis intersection fitting.
We employ the following fitting form~\cite{Jin:2017jjp} which incorporates the correction term
associated with the contribution of the energy-like observable,
\be
K
=
\left[
K_{\rm E}
+
AN_{\rm s}^{1/\nu}(\beta-\beta_{\rm E})
\right]
(1+BN_{\rm s}^{y_{\rm t}-y_{\rm h}}),
\label{eqn:new}
\ee
where $K_{\rm E}$, $\beta_{\rm E}$, $A$, $\nu$, $B$, and $y_{\rm t}-y_{\rm h}$ are basically fitting parameters.
Following Ref.~\cite{Jin:2017jjp}, we assume the three-dimensional Z$_2$ universality class for $K_{\rm E}=-1.369$, $\nu=0.630$, and $y_{\rm t}-y_{\rm h}=-0.894$, namely
the actual fitting parameters are now $\beta_{\rm E}$, $A$ and $B$.
The fitting results are shown in Table~\ref{tab:CEP} together with the previous smaller $N_{\rm t}\le10$ results.
The quality of the fitting for $N_{\rm t}=12$ is reasonable.
In the table, $\kappa_{\rm E}$ is estimated by an interpolated transition line in Eq.~(\ref{eqn:kappa_t})
together with the corresponding $\beta_{\rm E}$ as an input.

\begin{table}[t!]
\begin{ruledtabular}
\caption{\label{tab:CEP}
Fit results for kurtosis intersection with the fitting form in Eq.~(\ref{eqn:new}) for $N_{\rm t}=4-12$.
In the fitting we assume the 3D Z$_2$ universality class, namely
$K_{\rm E}=-1.396$, $\nu=0.630$ and $y_{\rm t}-y_{\rm h}=-0.894$ are fixed in the fitting procedure.
Using the central value of $\beta_{\rm E}$ as an input,
$\kappa_{\rm E}$ is obtained from the interpolation formula of the transition line in Eq.~(\ref{eqn:kappa_t}).
The error of $\kappa_{\rm E}$ contains that from only the interpolation procedure but not the error of $\beta_{\rm E}$.
}
\begin{tabular}{r|llllr}
$N_{\rm t}$
&
$\beta_{\rm E}$
&
$\kappa_{\rm E}$
&
$A$
&
$B$
&
$\chi^2/{\rm d.o.f.}$
\\
\hline
4&
$1.6099 ( 17 )$&$0.1430048 ( 13 )$&$0.311 ( 14 )$&$\phantom{-}0.10 ( 21 )$&$3.77$\\
6&
$1.72462 ( 40 )$&$0.1406334 ( 14 )$&$0.422 ( 12 )$&$-0.052 ( 52 )$&$0.70$\\
8&
$1.74953 ( 33 )$&$0.1402512 ( 10 )$&$0.414 ( 13 )$&$-1.33 ( 15 )$&$0.73$\\
10&
$1.77545 ( 53 )$&$0.1397274 ( 17 )$&$0.559 ( 29 )$&$-2.97 ( 25 )$&$0.43$\\
12&
$1.8105 ( 40 )$&$0.1390230 ( 16 )$&$0.41 ( 13 )$&$-7.4 ( 2.3 )$&$1.20$\\
\end{tabular}
\end{ruledtabular}
\end{table}

\subsection{Analysis for exponent of susceptibility peak height}
\begin{figure}[t]
\begin{center}
\includegraphics[bb=0 0 237 236,width=.5\textwidth]{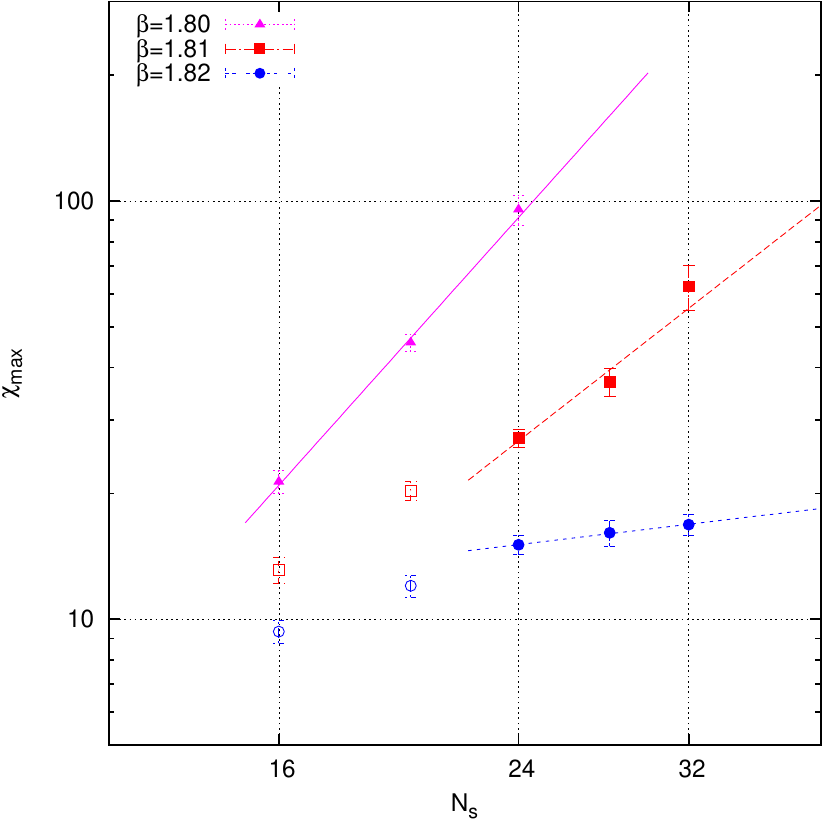}
\end{center}
\caption{
The volume scaling of susceptibility peak height for $N_{\rm t}=12$.
Both axes are scaled logarithmically.
The filled symbols are included in the fit but open ones are not
}
\label{fig:sus}
\end{figure}

\begin{figure}[t]
\begin{center}
\includegraphics[bb=0 0 353 168,width=.8\textwidth]{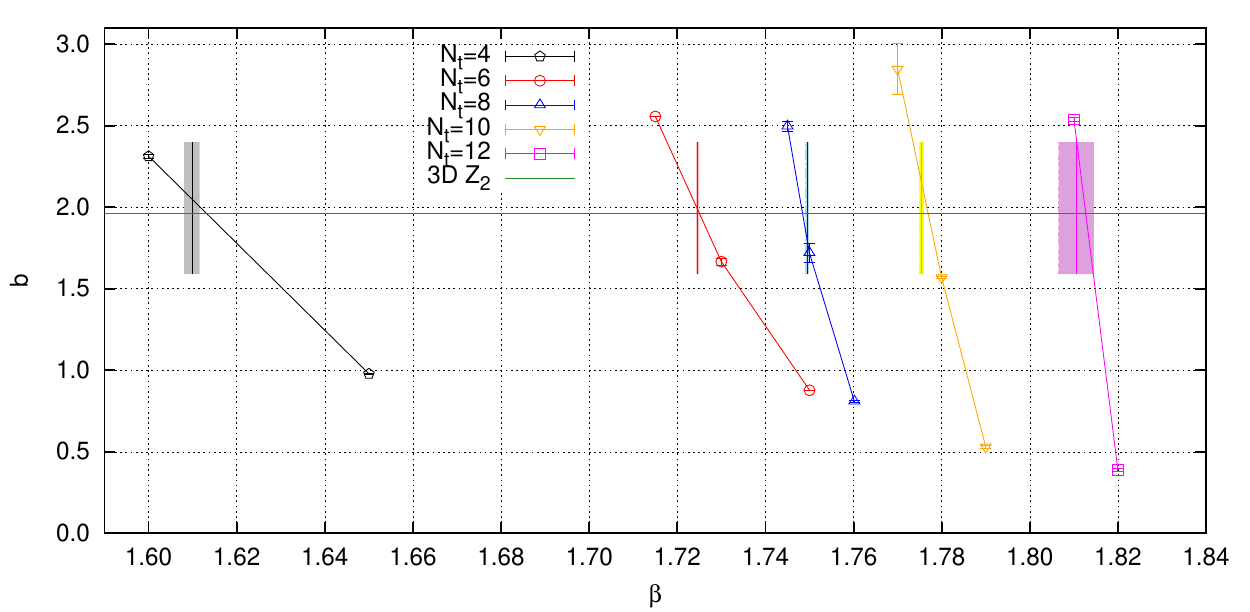}
\end{center}
\caption{
Exponent of the susceptibility peak height along the transition line projected on $\beta$ value for $N_{\rm t}=4$, $6$, $8$, $10$, and $12$.
The line connecting the data points is to guide readers' eyes.
The point where the line for each $N_{\rm t}$ intersects the (green) horizontal line is an estimate of the critical point assuming the Z$_2$ universality class.
On the other hand, the shaded areas represent the critical $\beta$ determined by the kurtosis intersection analysis.
}
\label{fig:exponent}
\end{figure}

As seen above, the kurtosis intersection analysis is not fully satisfactory since we have heavily relied on the assumption of the universality class of the critical point.
Therefore we should
cross check the location and the universality class of the critical point.
For that purpose, 
we investigate the scaling of the susceptibility peak height for the chiral condensate,
\begin{equation}
\chi_{\max}
\propto
(N_{\rm s})^b.
\label{eqn:chimax}
\end{equation}
At a critical point, the exponent should be $b=\gamma/\nu$, where
$\gamma$ and $\nu$ are critical exponents.
As discussed in Ref.~\cite{Jin:2017jjp},
in general there are correction terms in the above formula but here we neglect them just for a simplicity.
The data in Table~\ref{tab:kappat} is fitted with the above functional form as seen in Fig.~\ref{fig:sus}.
The resulting exponent $b$ is plotted in Fig.~\ref{fig:exponent} along the transition line projected on $\beta$.
Assuming the Z$_2$ universality class ($\gamma=1.237$ and $\nu=0.630$) provides an estimation of the critical point of $\beta$,
we confirm that it is consistent with that of the kurtosis intersection for $N_{\rm t}=12$ as well.
This cross check assures that our analysis is working well.

\subsection{Estimate of critical mass in continuum limit}
For scale setting, we perform the zero temperature simulations, which roughly cover the parameter range of the critical endpoints,
and their results are summarized in Appendix~\ref{sec:scale}.
For example, in Fig.~\ref{fig:hadron_kappa},
the pseudoscalar meson mass $m_{\rm PS}$ and the Wilson flow scale
$\sqrt{t_0}$~\cite{Luscher:2010iy} in lattice units are plotted as a function of $\kappa$ at $\beta=1.80$ and $1.81$.
The blue vertical line represents the location of the transition point $\kappa_{\rm t}(\beta,\infty)$ in Table~\ref{tab:kappat} for corresponding $\beta$.
From this figure, we obtain the hadronic quantity at the transition point.
Although the transition point is slightly out of the interpolation range, the monotonic behavior of data points suggests that such a short extrapolation should be harmless.

\begin{figure}[t!]
\begin{center}
\includegraphics[bb= 0 0 340 255, width=0.49\textwidth]{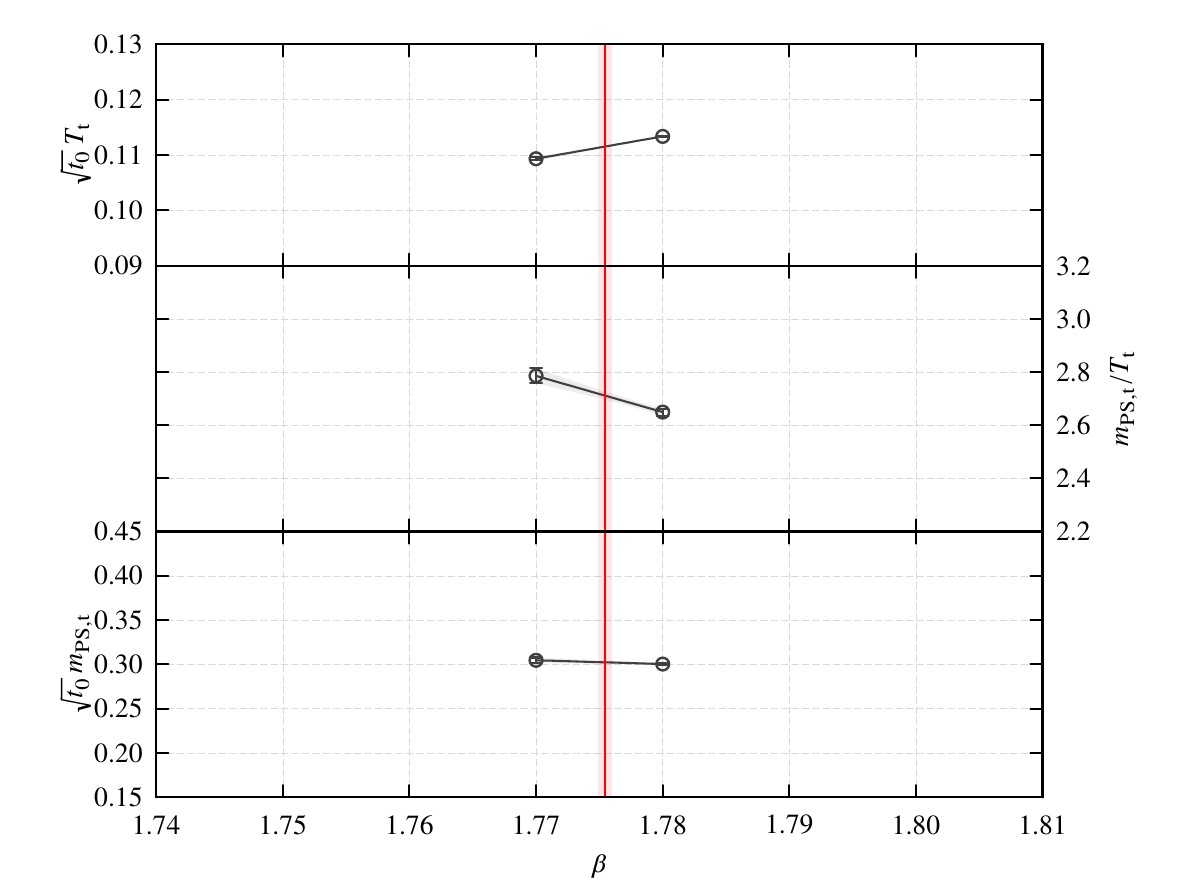}
\includegraphics[bb= 0 0 340 255, width=0.49\textwidth]{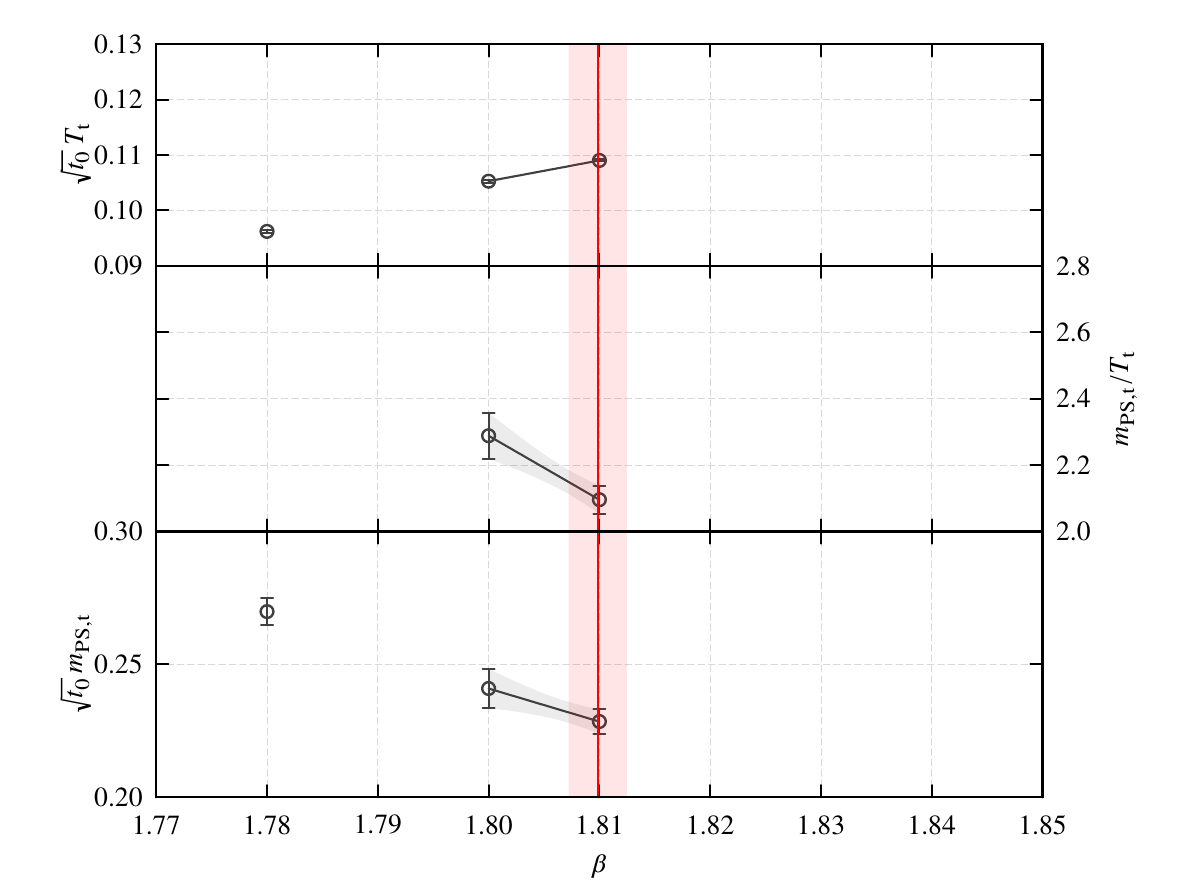}
\end{center}
\caption{
$\sqrt{t_0}T$, $m_{\rm PS}/T$, and $\sqrt{t_0}m_{\rm PS}$ along the transition line projected on $\beta$ for $N_{\rm t}=10$ (left)
and $12$ (right).
The vertical red line shows the location of the critical value of $\beta$ determined by the kurtosis intersection analysis.
}
\label{fig:hadron_transition_line}
\end{figure}

The dimensionless combination of the hadronic quantities $\sqrt{t_0}T$, $m_{\rm PS}/T$, and $\sqrt{t_0}m_{\rm PS}$ along the transition line projected on $\beta$
are plotted in Fig.~\ref{fig:hadron_transition_line} for $N_{\rm t}=10$ and $12$.
The vertical red line represents the location of the critical point determined by the kurtosis intersection method, and the plot allows us
to obtain the hadronic quantities at the critical point.
From an interpolation, one can obtain the critical value of the dimensionless quantities for each temporal size $N_{\rm t}$.
The actual numbers are summarized in Table~\ref{tab:lattice}.

\begin{figure}[h!]
\begin{center}
\includegraphics[bb=0 0 340 255,width=.45\textwidth]{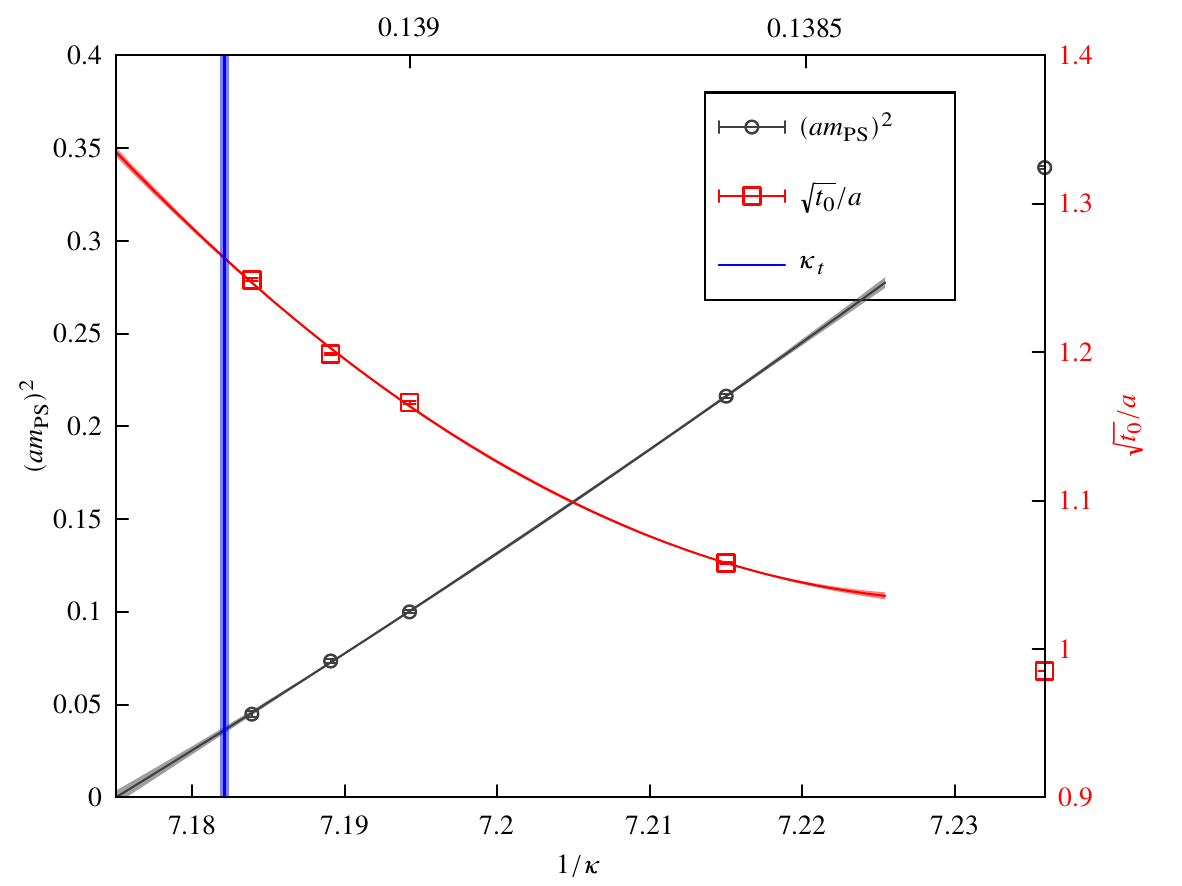}
\includegraphics[bb=0 0 340 255,width=.45\textwidth]{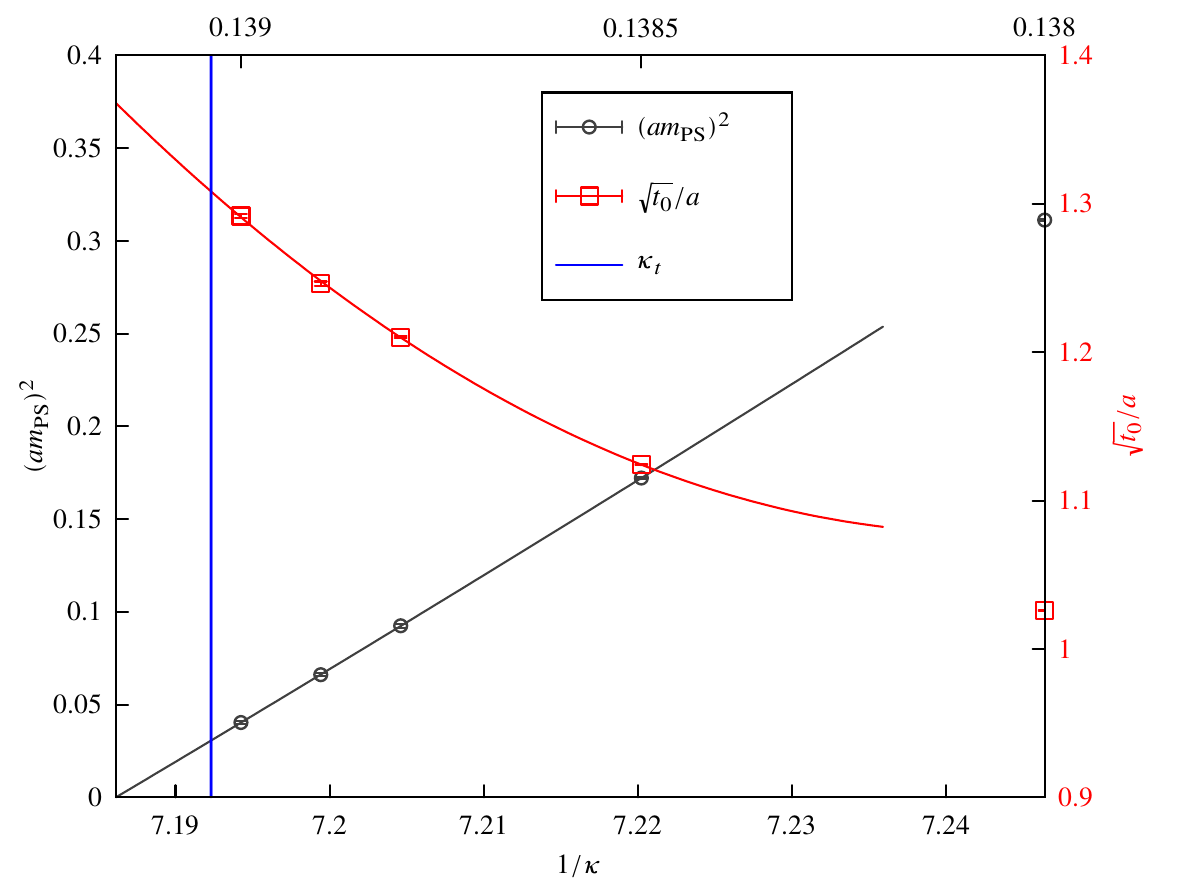}
\end{center}
\caption{
The hadronic quantities in lattice units $(am_{\rm PS})^2$, $\sqrt{t_0}/a$ as a function of $\kappa$ at $\beta=1.80$ (left) and $1.81$ (right).
The vertical blue line shows the location of the transition point for $\kappa$ at the corresponding $\beta$ with $N_{\rm t}=12$.}
\label{fig:hadron_kappa}
\end{figure}

\begin{table}[t!]
\caption{\label{tab:lattice}
The hadronic dimensionless quantities at the critical endpoint for $N_{\rm t}=4$, $6$, $8$, $10$, and 12.
Note that $N_{\rm t}=10$ results are updated compared with the previous work~\cite{Jin:2017jjp} since the hadronic quantities
at $\beta=1.78$ are updated as shown in Table~\ref{tab:scalefit}.
}
\begin{ruledtabular}
\begin{tabular}{l|lll}
$N_{\rm t}$
&
$\sqrt{t_0}m_{\rm PS,E}$
&
$\sqrt{t_0}T_{\rm E}$
&
$m_{\rm PS,E}/T_{\rm E}$
\\
\hline
 4 &      \phantom{-}0.6545(24) &     0.16409(13) &      \phantom{-}3.987(12)\\
 6 &      \phantom{-}0.5282(12) &     0.13328(23) &      \phantom{-}3.9630(63)\\
 8 &      \phantom{-}0.3977(19) &     0.11845(20) &      \phantom{-}3.357(16)\\
10 &      \phantom{-}0.3023(17) &     0.11154(26) &      \phantom{-}2.711(17)\\
12 &      \phantom{-}0.2287(57) &     0.1090(10)  &      \phantom{-}2.099(66)\\
\end{tabular}
\end{ruledtabular}
\end{table}

\begin{table}[t!]
\caption{\label{tab:continuum}
The continuum extrapolation of
the hadronic dimensionless quantities at the critical endpoint
with various functional forms and fitting ranges.
}
\begin{ruledtabular}
\begin{tabular}{lll|llllll}
name
&
functional form
&
fitting range
&
$\sqrt{t_0}m_{\rm PS,E}$
&
$\chi^2/\dof$
&
$\sqrt{t_0}T_{\rm E}$
&
$\chi^2/\dof$
&
$m_{\rm PS,E}/T_{\rm E}$
&
$\chi^2/\dof$
\\
\hline
A (fit)  & $a_0+a_1/N_{\rm t}^2$                & $N_{\rm t}=8$-$12$ &   \phantom{$-$}0.1243(52) & $12.07$ &  0.09943(34) & $0.63$ &   \phantom{$-$}1.491(50) & $13.64$\\
B (fit)  & $a_0+a_1/N_{\rm t}+a_2/N_{\rm t}^2$  & $N_{\rm t}=6$-$12$ &            $-$0.215(30)   & $\phantom{1}0.18$  &           -- &      --&            $-$2.12(29)   & $\phantom{1}0.72$\\
C (solve)& $a_0+a_1/N_{\rm t}^2$                & $N_{\rm t}=10$-$12$& \phantom{$-$}0.061(19)  &     --  &           -- &      --&   \phantom{$-$}0.71(22)  &    -- \\
D (solve)& $a_0+a_1/N_{\rm t}+a_2/N_{\rm t}^2$  & $N_{\rm t}=8$-$12$  &          $-$0.26(12)    &     --  &           -- &      --&            $-$3.2(1.3)   &    -- \\
\end{tabular}
\end{ruledtabular}
\end{table}

\begin{figure}[t]
\begin{center}
\includegraphics[bb= 0 0 340 255, width=0.45\textwidth]{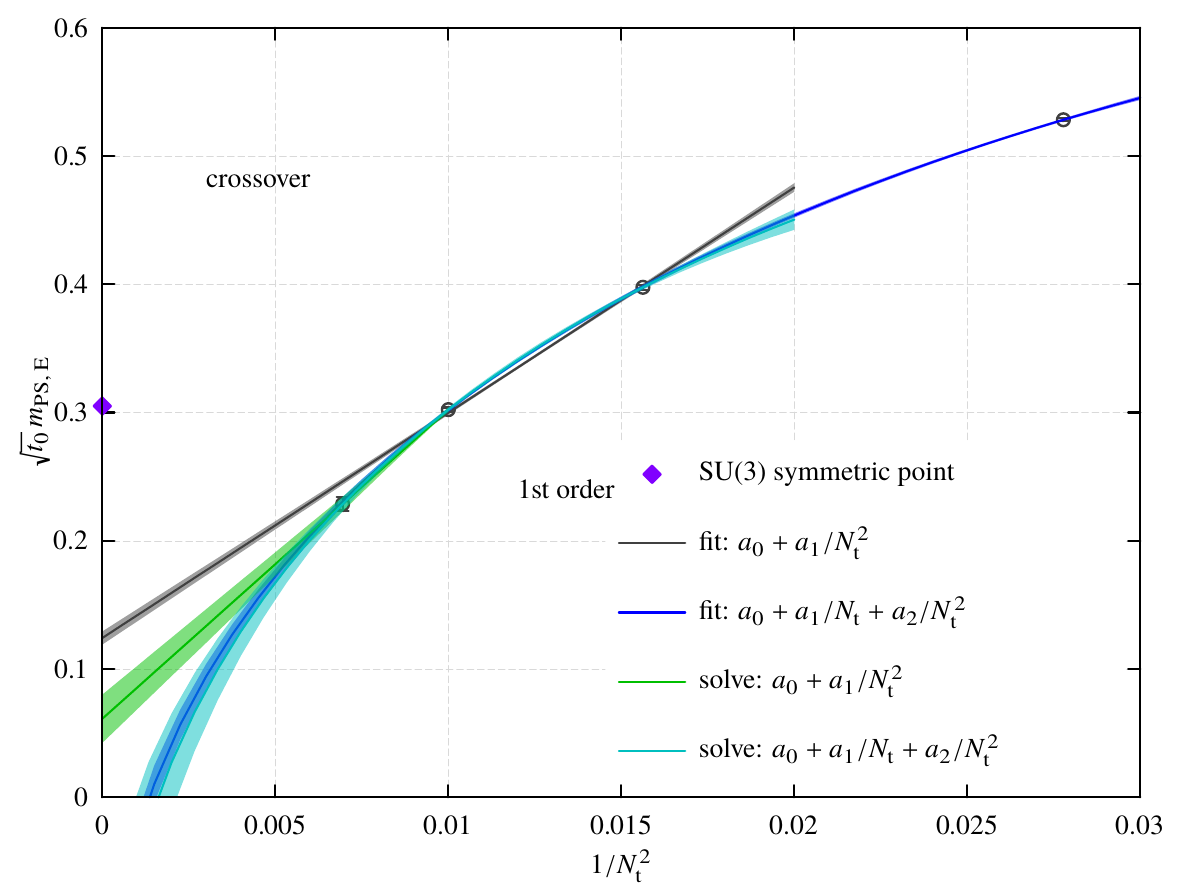}
\includegraphics[bb= 0 0 340 255, width=0.45\textwidth]{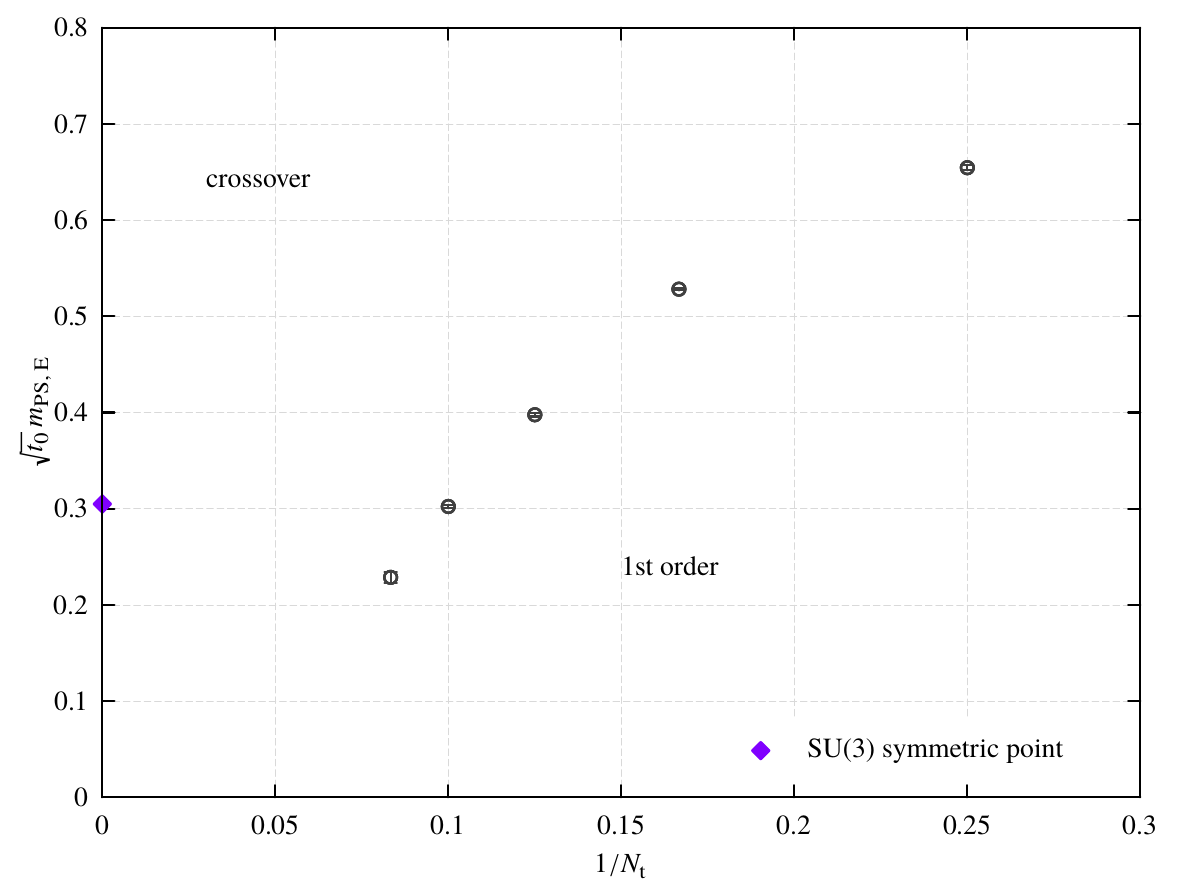}
\includegraphics[bb= 0 0 340 255, width=0.45\textwidth]{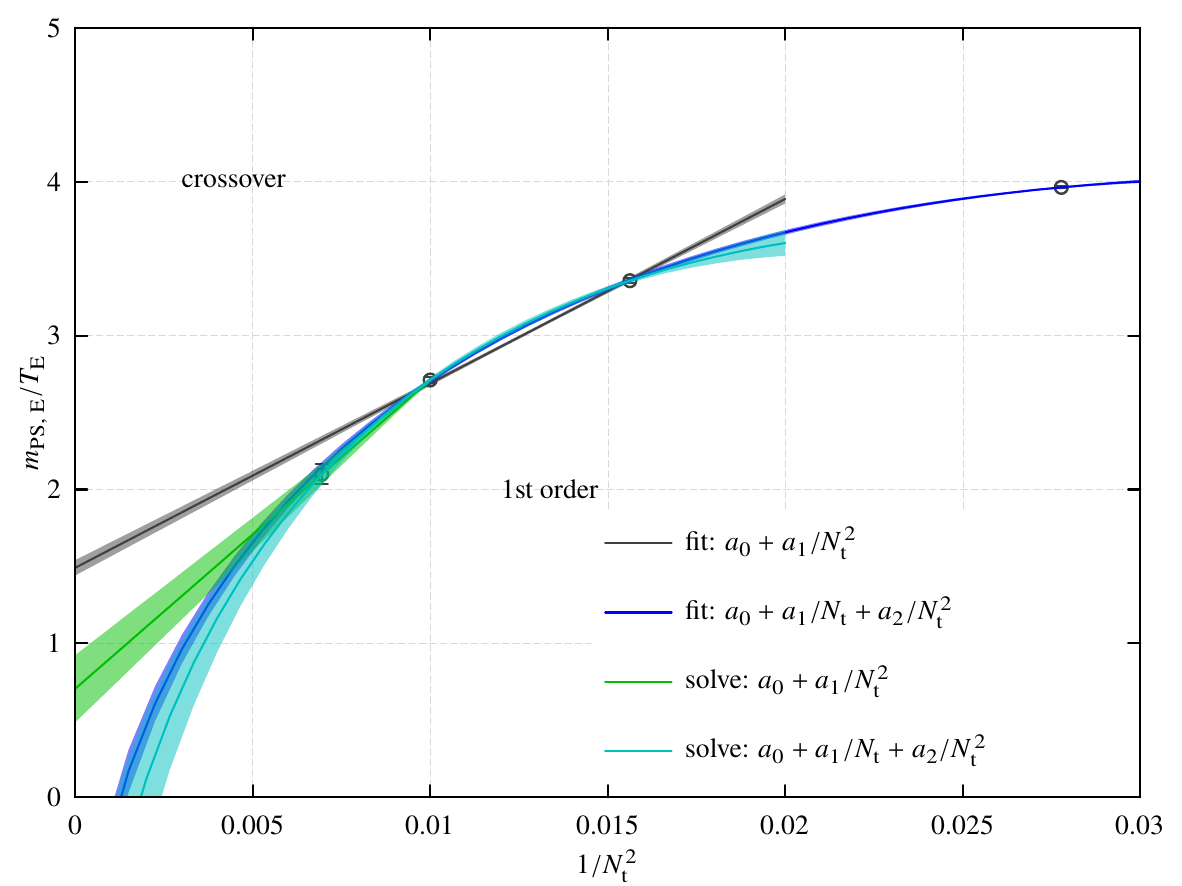}
\includegraphics[bb= 0 0 340 255, width=0.45\textwidth]{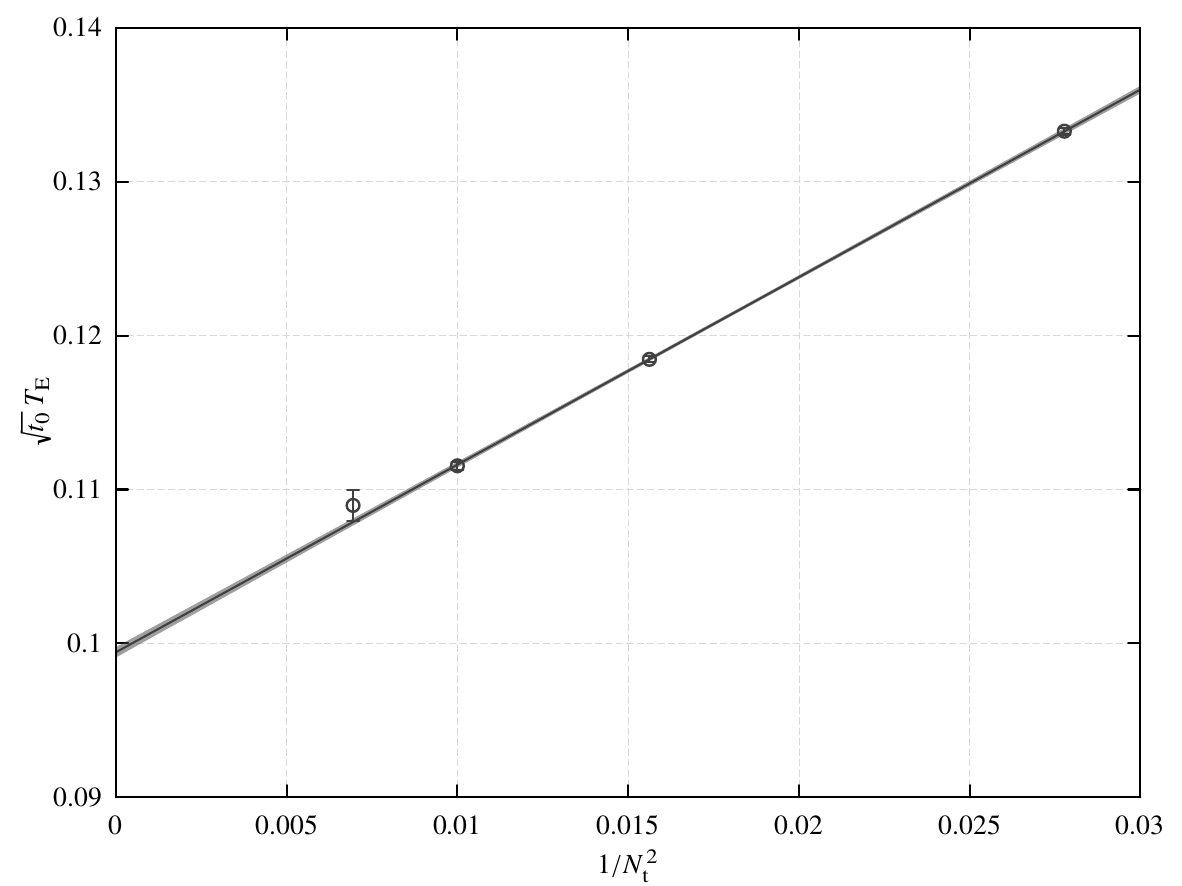}
\end{center}
\caption{
Continuum extrapolation of the critical endpoint.
In upper panels, the left one is for $\sqrt{t_0}m_{\rm PS,E}$ vs $1/N_{\rm t}^2$ while
the right one is for $\sqrt{t_0}m_{\rm PS,E}$ vs $1/N_{\rm t}$.
In the lower panel
$m_{\rm PS,E}/T_{\rm E}$
(left)
and
$\sqrt{t_0}T_{\rm E}$
(right) are plotted as a function of $1/N_{\rm t}^2$.
}
\label{fig:continuum}
\end{figure}

\begin{figure}[hbt!]
\begin{center}
\includegraphics[bb= 0 0 255 340, width=0.4\textwidth]{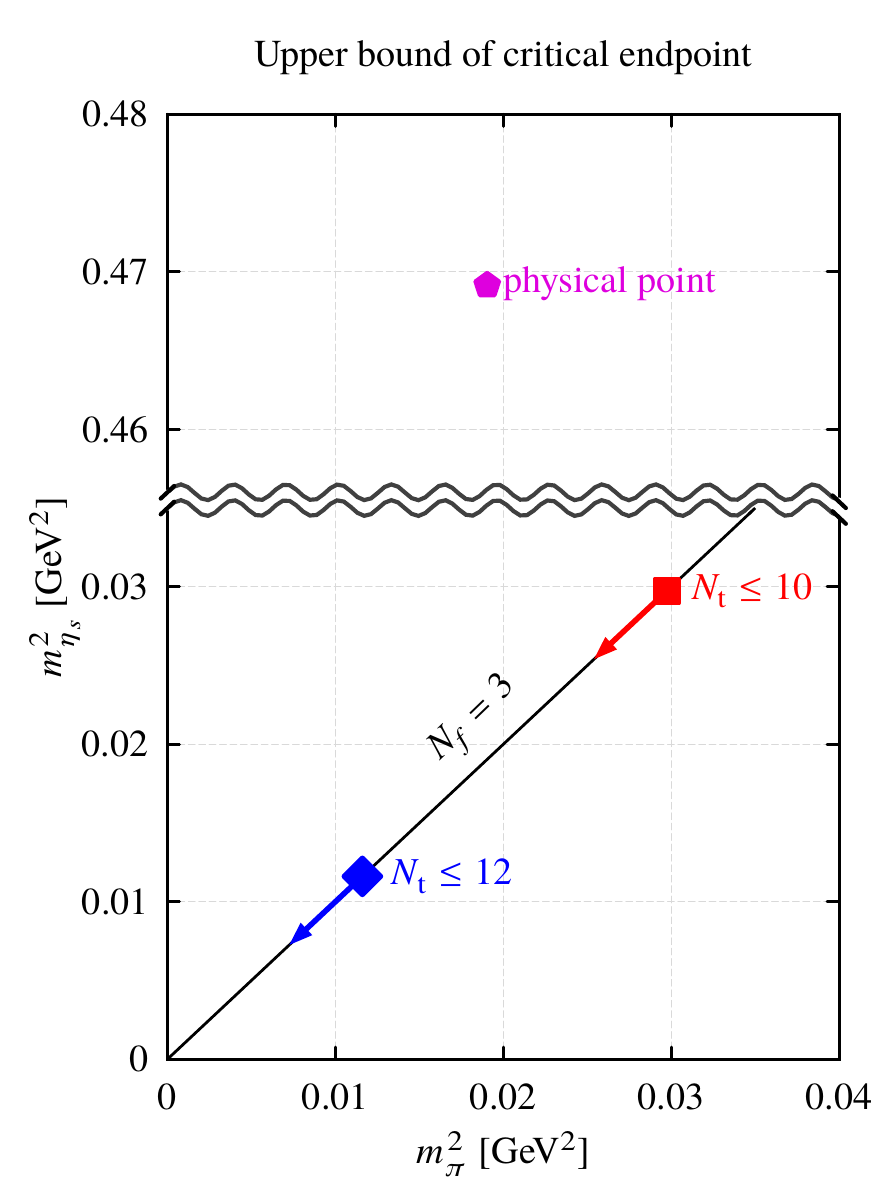}
\end{center}
\caption{Columbia-like plot with axes $m_\pi^2$ and $m_{\eta_s}^2$ in physical units.
The blue symbol denotes the upper bound obtained in this work $N_{\rm t}\le12$ while
the red one is given in our previous study $N_{\rm t}\le 10$ \cite{Jin:2017jjp}.
The physical point is also shown just for a reference.
}
\label{fig:CP1}
\end{figure}

Figure~\ref{fig:continuum} shows the continuum extrapolation of $\sqrt{t_0}m_{\rm PS,E}$, $m_{\rm PS,E}/T_{\rm E}$, and $\sqrt{t_0}T_{\rm E}$.
As for $\sqrt{t_0}T_{\rm E}$ (lower right panel of Fig.~\ref{fig:continuum}), even though the new data point at $N_{\rm t}=12$ is included,
a stable continuum extrapolation is observed and we obtain $\sqrt{t_0}T_{\rm E}=0.09943(34)$
which has no significant difference compared with the previous one $\sqrt{t_0}T_{\rm E}=0.09970(37)$ in Ref.~\cite{Jin:2017jjp}.
In terms of the physical units the critical temperature is given by
$T_{\rm E}=134(3)$~MeV,
where we have used the Wilson flow scale $1/\sqrt{t_0}=1.347(30)$~GeV in Ref.~\cite{Borsanyi:2012zs} as an input.
On the other hand, in Fig.~\ref{fig:continuum}, $\sqrt{t_0}m_{\rm PS,E}$ and $m_{\rm PS,E}/T_{\rm E}$ show significantly large scaling violation.
In the extrapolation procedure, we try some functional forms including up to quadratic correction term and examine the fitting range dependence.
As a result, their dependence turns out to be quite large as shown in Fig.~\ref{fig:continuum} (upper left for $\sqrt{t_0}m_{\rm PS,E}$
and lower left for $m_{\rm PS,E}/T_{\rm E}$) and Table~\ref{tab:continuum}.
The fitting named as (A) is not acceptable since the $\chi^2/\dof$ is very large.
For other cases, $\chi^2/\dof$ is reasonable but in some cases, the extrapolated mass is negative.
Furthermore we plot $\sqrt{t_0}m_{\rm PS,E}$ as a function of $1/N_{\rm t}$ in Fig.~\ref{fig:continuum} (upper right),
which shows a linear scaling behavior thus the leading scaling violation seems O($a$) for this quantity.
Since our value of $c_{\rm sw}$ around $\beta\sim1.81$ is out of the interpolation range\footnote{
In ref~\cite{Aoki:2005et}, the constant physics condition is used to determine $c_{\rm sw}$.
Actually this condition makes the determination of $c_{\rm sw}$ at low $\beta$ very hard,
since one needs extremely small lattice size $N_{\rm s}\ll6$ for such a low $\beta$ case
in order to keep the physical length scale constant.
Of course one can change the constant physics condition such that a low $\beta$ simulation
is feasible with a reasonable lattice size say $N_{\rm s}=6$,
but in that case the physical lattice size is larger and then a high $\beta$ simulation requires quite large lattice size $N_{\rm s}\gg6$
and moreover infrared cutoff thanks to the boundary condition gets weaker.
} ($\beta\ge1.90$),
it is likely that the O($a$) improvement program does not work well in our parameter range.
Thus to avoid such an extrapolation, in the future we should do large $\beta>1.90$ simulations, that is,
very large $N_{\rm t}$ simulations where the O($a$) improvement works well.
In such a simulation, O($a^2$) scaling may be seen and an extrapolation to the negative value could be avoided.
In any case, here we conservatively quote an upper bound of the critical value
$\sqrt{t_0}m_{\rm PS,E}\lesssim0.08$, which is taken from the maximum continuum value among all the fits except for (A).
In physical units, this bound is
$m_{\rm PS,E}\lesssim 110$~MeV,
which is smaller than our previous estimate ($m_{\rm PS,E}\lesssim$170~MeV)~\cite{Jin:2017jjp}.
A Columbia-like plot, whose axes are given by hadron masses, is shown in Fig.~\ref{fig:CP1} to
display the current situation of our study.
For future references, we mention the continuum extrapolation of $m_{\rm PS,E}/T_{\rm E}$
in Fig.~\ref{fig:continuum} (lower left) where large cutoff dependence is seen as well.
Thus we quote an upper bound
$m_{\rm PS,E}/T_{\rm E}\lesssim0.93$ obtained with the same criteria as that of
$\sqrt{t_0}m_{\rm PS,E}$.

\section{Summary and outlook}
\label{sec:summary}

In this study, we performed the large scale simulations for $N_{\rm t}=12$ by using the Wilson-type fermions.
This is an extension of our previous works at the smaller temporal size simulations $N_{\rm t}\le10$~\cite{Jin:2014hea,Jin:2017jjp}.
By using the modified formula of the kurtosis intersection analysis, the critical endpoint is determined with assuming 3D Z$_2$ universality class.
The continuum limit for the critical temperature is smoothly taken and we obtain
$T_{\rm E}=134(3)$~MeV
which is essentially the same as before.
On the other hand, for the critical mass, the continuum extrapolation significantly dominates the systematic error, thus here we conservatively quote upper bounds
\begin{eqnarray}
m_{\rm PS,E}
\lesssim
110\,\,
{\rm MeV},
\,\,\,\,\,\,
m_{\rm PS,E}/T_{\rm E}
\lesssim
0.93,
\,\,\,\,\,\,
\end{eqnarray}
where we have made the upper bound about $40\%$ smaller than before.

In fact, the studies using the staggered-type fermions suggested much lower bound
$m_{\rm PS,E}\lesssim50$~MeV in Ref.~\cite{Bazavov:2017xul}.
Thus it is likely that the critical mass is so small that modern computers cannot access it directly, or it could be zero.
Moreover an insightful result for $N_{\rm f}=4$ QCD was reported by de Forcrand and D'Elia~\cite{deForcrand:2017cgb}
where the standard staggered fermions are used to study the critical point.
They found large cutoff effects compared with $N_{\rm f}=3$ case and
the critical mass tends to be zero with decreasing the lattice spacing.
A similar tendency is observed even in the Wilson-type fermions by our group~\cite{Ohno:2018gcx}.
Since there is no rooting issue when the number of flavor is a multiple of $4$,
the feature that the critical mass is extremely small for multiple-flavor QCD seems to be robust.

Of course, in order to make a quantitative conclusion, one has to carry out large $N_{\rm t}$ simulations or use the improved lattice actions.
Another possibility is to invent a new analysis method which is useful
to study such a near-zero critical mass.

\section*{Acknowledgements}
This research was supported by Multidisciplinary Cooperative Research Program in CCS, University of Tsukuba
and projects of the RIKEN Supercomputer System.

\appendix

\section{Wilson flow scale and pseudoscalar meson mass at zero temperature}
\label{sec:scale}

Simulation parameters, results for the pseudoscalar meson mass $am_{\rm PS}$, and Wilson flow scale parameter $\sqrt{t_0}/a$
 are summarized in Table~\ref{tab:scale}. 
Result of following combined fit is given in Table~\ref{tab:scalefit},
\begin{eqnarray}
\label{eq:scalefit-1}
 (am_{\rm PS})^2 &=& a_1 \left({1\over \kappa} - {1 \over \kappa_{\rm c}}\right) + a_2 \left({1\over \kappa} - {1\over \kappa_{\rm c}}\right)^2 \,, \\
 \label{eq:scalefit-2}
{\sqrt{t_0} \over a} &=& b_0 + b_1 \left({1\over \kappa} - {1 \over \kappa_{\rm c}}\right) + b_2 \left({1\over \kappa} - {1 \over \kappa_{\rm c}}\right)^2\,.
\end{eqnarray}

\begin{table}[h]
\caption{\label{tab:scale}
Simulation parameters $\kappa$, $N_{\rm s}$, and $N_{\rm t}$ and measured$\sqrt{t_0}/a$ and $am_{\rm PS}$ at $\beta=1.77$, $1.78$, $1.80$, and $1.81$.
Note that the data at $\beta=1.78$ is updated compared with the previous work~\cite{Jin:2017jjp}.
}
\begin{ruledtabular}
\begin{tabular}{cccc|ll}
$\beta$&$\kappa$ & $N_{\rm s}$ & $N_{\rm t}$ & $\sqrt{t_0}/a$ &  $am_{\rm PS}$  \\
\colrule
1.77&
  0.137100 &   12 &   24 &     0.77014(39) &      1.0040(12) \\
 & 0.137670 &   12 &   24 &     0.79076(35) &     0.91999(86) \\
 & 0.138500 &   12 &   24 &     0.83773(53) &      0.7675(12) \\
 & 0.138700 &   12 &   24 &     0.85652(79) &      0.7172(18) \\
 & 0.138903 &   16 &   32 &     0.87524(52) &     0.66902(80) \\
 & 0.139000 &   16 &   32 &     0.88795(57) &     0.63966(81) \\
 & 0.139653 &   16 &   32 &      1.0096(13) &      0.4063(14) \\
 & 0.139750 &   16 &   32 &      1.0447(17) &      0.3528(20) \\
 & 0.139850 &   16 &   32 &      1.0903(34) &      0.2851(36) \\
 & 0.139900 &   16 &   32 &      1.1163(52) &      0.2433(49) \\
\colrule
1.78&
  0.139356 &   16 &   32 &      1.0299(67) &      0.4057(18) \\
&  0.139500 &   24 &   48 &     1.08443(70) &     0.33595(69) \\
&  0.139600 &   24 &   48 &     1.12526(92) &     0.27661(79) \\
&  0.139650 &   24 &   48 &      1.1505(12) &      0.2399(14) \\
&  0.139700 &   24 &   48 &      1.1808(14) &      0.1922(12) \\
\colrule
1.80&
  0.138200 &   32 &   64 &     0.98521(21) &     0.58267(65) \\
&  0.138600 &   32 &   64 &     1.05792(35) &     0.46505(96) \\
&  0.139000 &   32 &   64 &      1.1662(11) &      0.3162(13) \\
&  0.139100 &   32 &   64 &     1.19878(98) &      0.2711(19) \\
&  0.139200 &   32 &   64 &      1.2487(11) &      0.2118(36) \\
\colrule
1.81&
  0.138000 &   32 &   64 &     1.02589(23) &     0.55788(58) \\
&  0.138500 &   32 &   64 &     1.12419(37) &     0.41466(78) \\
&  0.138800 &   32 &   64 &     1.21006(60) &      0.3042(17) \\
&  0.138900 &   32 &   64 &      1.2462(15) &      0.2572(15) \\
&  0.139000 &   32 &   64 &      1.2917(14) &      0.2006(21) \\
\end{tabular}
\end{ruledtabular}
\end{table}

\begin{table}[h]
\caption{\label{tab:scalefit}
Fit results to Eqs.~(\ref{eq:scalefit-1}) and~(\ref{eq:scalefit-2})  for the critical hopping parameter $\kappa_{\rm c}$ and 
coefficients for the pseudoscalar meson mass $am_{\rm PS}$ and the Wilson flow parameter $\sqrt{t_0}/a$ for  $\beta=1.77$, $1.78$, $1.80$, and $1.81$.
Note that the data at $\beta=1.78$ is updated compared with the previous work~\cite{Jin:2017jjp}.
}
\begin{ruledtabular}
\begin{tabular}{c|ccccccc|c}
$\beta$ & $\kappa_{\rm c}$ & $a_1$ & $a_2$ &$b_0$ &  $b_1$ & $b_2$ & $\chi^2/\dof$  & fit range \\
            &                 &          &         &        &            &         &                  &  $\kappa>$  \\
\colrule
 1.77 &        0.1400313(72) &        9.00(27) &      $-$24.2(4.2) &      1.1814(69) &      $-$10.85(46) &      100.3(6.5) &       0.34 & 0.1390 \\
 1.78 &        0.1397915(29) &        8.12(23) &         $-$36(11) &      1.2380(43) &      $-$13.51(76) &         214(35) &       2.51 & 0.1390 \\
 1.80 &         0.139374(14) &        5.02(34) &        9.6(6.0) &      1.3353(82) &      $-$10.74(33) &       95.2(5.6) &      13.88 & 0.1384 \\
 1.81 &        0.1391557(99) &        4.98(29) &        2.3(6.4) &      1.3673(62) &      $-$10.21(33) &       90.0(6.4) &       1.02 & 0.1382 \\
\end{tabular}
\end{ruledtabular}
\end{table}

\end{document}